# A Bragg glass phase in the vortex lattice of a type II superconductor

T. Klein*, I. Joumard*, S. Blanchard*, J. Marcus*, R. Cubitt†, T. Giamarchi‡ & P. Le Doussal§

*Laboratoire d'Etudes des Propriétés Electroniques des Solide - CNRS, BP166, 38042 Grenoble Cedex 9, France

†Institut Laue Langevin, BP 156, 38042 Grenoble Cedex 9, France

‡Laboratoire de Physique des Solides, CNRS-UMR8502, UPS Bât 510, 91405 Orsay, France

§CNRS-Laboratoire de Physique Théorique de l'Ecole Normale Supérieure, 24 rue Lhomond, 75231 Paris, France

**Although crystals are usually quite stable, they are sensitive to a disordered environment: even an infinitesimal amount of impurities can lead to the destruction of the crystalline order[1]. The resulting state of matter has been a longstanding puzzle. Until recently it was believed to be an amorphous state in which the crystal would break into "crystallites"[2]. But a different theory[3] predicts the existence of a novel phase of matter: the so-called Bragg glass, which is a glass and yet nearly as ordered as a perfect crystal. The 'lattice' of vortices that can contain magnetic flux in type II superconductors provide a good system to investigate these ideas[4]. Here we show that neutron diffraction data of the vortex lattice in type II superconductors provides unambiguous evidence for a weak, power-law decay of the crystalline order characteristic of a Bragg glass. The theory also predicts accurately the electrical transport properties of superconductors; it naturally explains the observed phase transitions[4,5,6] and the dramatic jumps in the critical current[7,8] associated with the melting of the Bragg**

**glass. Moreover the model explains experiments as diverse as X-ray scattering in disordered liquid crystals[9,10] and conductivity of electronic crystals[11,12].**

After the Bragg glass was first proposed[3] its existence was supported by further analytical[13,14,15] and numerical[16,17] calculations. However, up to now experimental evidence for this phase has been indirect. For instance, the presence of a Bragg glass existing as a stable thermodynamic phase is expected to impose a universal phase diagram for type II superconductors[3,18,19] - and similar experimental phase diagrams have indeed been observed in a large variety of compounds[4,5,6]. Moreover, recent transport measurements in BiSrCaCuO[20] are consistent with the Bragg glass predictions[21,3]. But given the complexity of these materials it is quite difficult to rule out other interpretations of the data. To decide unambiguously on the nature of the vortex solid, it is necessary to use an imaging technique able to probe *directly* the topology of the system. Even if decoration and imaging experiments show large regions free of dislocations[22] or well ordered structures when the lattice is set into motion[23,24], these regions are still too small to provide an unambiguous signature of the Bragg glass: the expected power law divergence of the Bragg peaks. Diffraction experiments such as Small Angle Neutron Scattering (SANS) [25,26,27,28] can be performed at larger magnetic fields, and can thus bring the scale at which disorder becomes relevant down to the experimental resolution. When properly analyzed SANS experiments are thus an ideal method for deciding on this issue.

The experiments we report here were performed on a large single phased $(K,Ba)BiO_3$ crystal (mass ≈ 300mg, $T_c$ ≈ 23K). This compound has the advantage of being totally isotropic enabling us to avoid complications due to anisotropy such as a possible 3D-2D crossover. Our experiments were performed on the D11 line at the Institut Laue Langevin - Grenoble. Neutrons are coherently diffracted by the flux lines when the Bragg condition is fulfilled. In order to study the positional order of the vortex lattice, it is necessary to measure the correlation functions perpendicular to the vortices (that is in the detector plane). However, the poor experimental in-plane resolution (a few lattice constants) does not allow to probe these correlations directly. The solution is to measure the correlations *along* the vortex lines which can, on the contrary, be probed with good accuracy and are directly related to the in-plane correlations through the elastic constants. The sample is thus rocked through the Bragg angle (by an angle ω) and the diffracted intensity is integrated in the detector plane, giving the so-called rocking curve. The scattering geometry is shown in the top panel of Fig.1. In the vicinity of a wavevector $K_0$ the integrated intensity is then related to the structure factor S(q) through : $I(\omega) = F^2 \int dq_x \int dq_y S(q_x,q_y,K_0\omega)$ where F is the standard form factor of a single vortex[29]. The rocking curves obtained at 2K for $K_0 \approx 3^{1/2}/2.2\pi/a_0$, where $a_0$ is the vortex lattice spacing are shown in the lower panel of Fig. 1. The most direct way to check for the existence of the Bragg glass would be to deconvolute the experimental resolution and directly check the predicted[3] power law decrease of S(q). In superconductors, the intensity in the rocking curve tails is too weak to do this reliably. Fortunately as we will show, this power law decay can also be obtained from the *magnetic field dependence* of the rocking curves which turns out to provides an unambiguous evidence for the existence of the Bragg glass.

As shown in Fig.1, $I/F^2a_0$ is field independent below a magnetic field H*~ 0.7T but rapidly collapses at higher fields. However, despite this strong reduction of the intensity at the Bragg angle (i.e. of $S(q_z=0)$) the half width at half maximum of the rocking curves σ remains *magnetic field independent* ~ 0.18°; this value is close to the experimental resolution and remains constant up the highest magnetic field. This is a very striking result because the London theory predicts that the area under the rocking curve (Fig. 1) should remain roughly constant. A similar decrease of the diffracted intensity without any broadening of the rocking curve was also observed[25] in BiSrCaCuO and attributed to a 3D-2D crossover. This explanation however cannot hold in the perfectly isotropic $(K,Ba)BiO_3$ system.

How can this unusual behavior be understood? Let us first recall that for a perfect crystal, the finite experimental resolution (in momentum $\Delta q \sim 1/\xi$) would transform the ideal delta-function peak of the structure factor at $K_0$ into a broadened peak of height $\xi^d$ and half width $1/\xi$, (where d is the spatial dimension). If the positional order in the system is not perfect but decays over the characteristic length scale $R_a$ (for which displacements are of the order of the lattice spacing (that is $u(R_a) \sim a_0$)) the observed Bragg peak remains determined by the experimental resolution as long as $\xi < R_a$. However, the structure factor will become dependent on the decay of the translational order in the system for $\xi > R_a$. This will then allow us to distinguish between an amorphous system full of dislocations and the Bragg glass. In the first case, the translational order decays very rapidly beyond $R_a$ (for example exponentially with the distance) leading to a peak of height $R_a^d$ and half width in momentum $1/R_a$ (the influence of the experimental resolution is negligible in this case); a decrease in $R_a$ will thus lead to a decrease of the height of the peak *and* a broadening of the peak. The situation is very different in the Bragg glass. Indeed, in this case the positional order decays only weakly beyond $R_a$ (as a power-law of the distance with an exponent $\eta \sim 1$). The same power law decay is predicted[3] perpendicularly and parallel to the vortices and the corresponding characteristic lengths $R_{az}$ and $R_{axy}$ are connected through $R_{az} = R_{axy} \sqrt{c_{44}/c_{66}}$ (where the $c_{ii}$ are elastics constants). In the absence of any experimental limitation this would lead[3] to power-law divergent Bragg peaks for $q \to 0$: $S(q) \sim 1/q^{d-\eta}$. In a real experiment however, $\xi$ suppresses this divergence and the shape of the peak is then controlled *both* by $R_a$ (in the appropriate direction) and $\xi$.

Our calculation shows that, for $\xi > R_a$, the structure factor (and hence the rocking curve) has the form shown in Fig. 2. The most important point is that, for $d \geq \eta$, the width at half maximum of the peak remains entirely given by the experimental resolution $1/\xi$ for both $R_a > \xi$ and $R_a < \xi$ whereas the height of the peak $S(q_z=0)$ is proportional to $R_a^\eta$ for $R_a < \xi$. In the Bragg glass, a decrease in $R_a$ thus leads to a decrease of the diffraction peak *without any broadening* . As shown in Fig.1, this behavior is in excellent agreement with the experimental data for H > H* ~ 0.7T (for H < H* $R_a$ > $\xi$ and the intensity remains constant). We note that such an unusual collapse of the rocking curve is intrinsically related to the power law behavior of S(q) and could not be observed for any faster decay of the positional order. Moreover, because the rocking curves measure the diffraction along the z direction, one has to set d=1 in the above expressions and this behavior thus *imposes* that $\eta \approx 1 = d$ (the half width at half maximum $\sigma$ is again expected to increase as $a_0/R_a$ for $(\eta-d) \log(\xi/R_a) > 1$) in excellent agreement with the Bragg glass model[3] ($\eta$ ~ 1-1.2) as well as numerical simulations[17]. The collapse of the rocking curve without any broadening is thus direct proof of the existence of the Bragg glass.

In order to further test the consistency of this explanation, we have extracted the positional correlation length $R_{az}$ from the data, using the Bragg glass formula for the intensity ($I(\omega=0)/F^2 \sim R_{az}$). The unknown overall normalization of the intensity is fixed by the condition that $R_{az} \sim \xi = 1/(K_0 \sigma) \sim 50\ a_0$ when the intensity starts to depend on the magnetic field. The field dependence of $R_{az}$ is plotted in Fig. 3. Because, for H >> $H_{c1}$ (where $H_{c1}$ is the lower critical field) increasing the field is equivalent to increase the effective disorder in the system, the observed decrease of $I(\omega=0)/F^2$ is in good agreement with the expected decrease of $R_{az}$. The solid line in Fig. 3 corresponds to a $1/B^{3/2}$ variation for $R_{az}$, which is consistent with the calculated dependence of $R_{az}$ on the magnetic field B using a simple elastic description for the vortex lattice[19]. The intensity finally drops rapidly towards zero at a field $B_m$ lying close to the onset of the second peak in magnetization measurements, thus confirming the theoretical interpretation of the second peak in terms of Bragg glass 'melting'. Moreover, neutron data give a very reasonable value for the positional correlation length at the transition : $R_{az}$ (here of the order of $R_{axy}$) ~ 20 $a_0$. As shown on Fig.3, this interpretation can also be applied to BiSrCaCuO. Indeed, similar behavior is observed in the two systems provided that the field axis is renormalized by the field for which the diffracted intensity vanishes (this moreover shows that there is only one relevant parameter: $B_m$, which is directly related to $R_a(0)$ that is the residual amount of disorder in the system).

The Bragg glass theory also explains the puzzling behavior of the measured temperature dependence of the intensity I (Fig. 4). As shown, the $1/\lambda^4$ dependence of the intensity expected in the standard London model (I ~ $F^2$) only provides a very poor agreement with the experimental data using a classical two-fluid model for temperature dependence of the magnetic penetration length $\lambda(T)$ (dotted line). This discrepancy was also pointed out in YbaCuO (ref 26) and remained unexplained. The neutron data can be fitted by introducing a phenomenological modification of the temperature dependence of the penetration length. However, such a strong deviation from the classical model has not been observed in the $\lambda(T)$ deduced from muon spin resonance relaxation data[26]. This shows that besides being an *ad hoc* fit to the data, such a modification of $\lambda(T)$ does not in fact contain any physical meaning. As muons are mostly sensitive to the behavior of individual vortices, the deviation observed in neutron data can be directly related to the long range correlation in the vortex lattice.

Indeed, in the Bragg glass model, an additional temperature dependence arises from $R_a(T)$ which is expected to vary as $(L_c/\xi_0^2)^3$ in the elastic theory (where $\xi_0$ is the superconducting coherence length and $L_c$ the collective pinning length which describes the strength of pinning). Assuming that pinning is related to fluctuations in the critical temperature (the so-called $\delta$-$T_c$ pinning model) $L_c \propto \xi_0^{2/3}$ and the resulting temperature dependence is plotted on Fig. 4 (dashed line). As shown, this yields very reasonable agreement with the experimental data without any adjustable parameter.

**Acknowledgements:**

We thank C. Simon for many invaluable discussions.

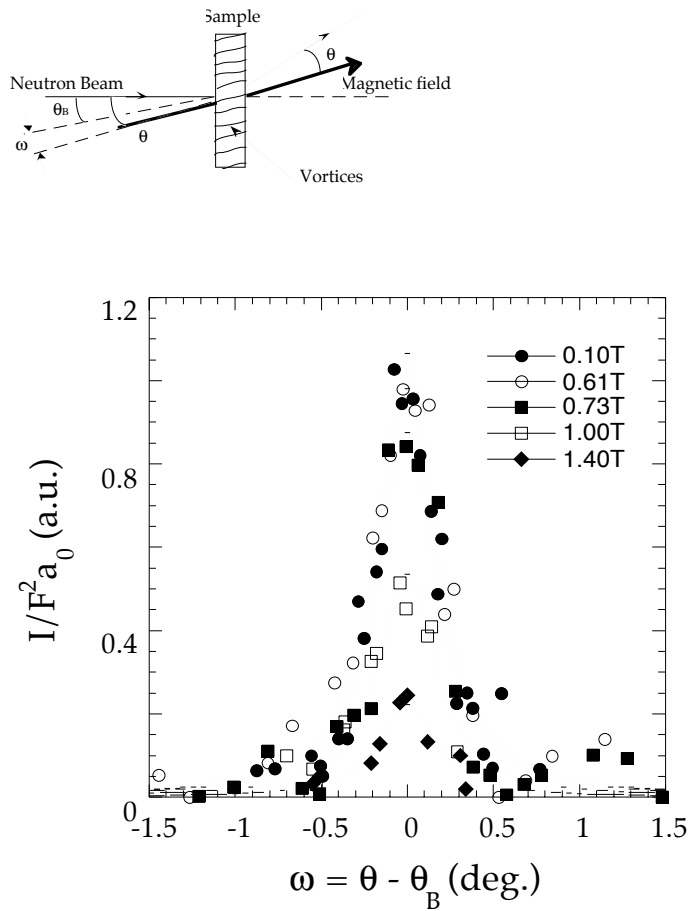

Figure 1: Angular dependence of neutron intensity diffracted by a $(K,Ba)BiO_3$ crystal and the sample geometry. Data were taken at 2K for different magnetic fields. The neutron beam (of wavelength 10A) was approximately parallel to the [111] direction of the crystal. The sample has been cooled in a slightly oscillating field (of the order of 5 % of the dc field) in order to obtain a good in plane orientational order. ω is the angle by which the sample is rocked away from the Bragg angle $θ_B$ (θ being the angle between the neutron beam and the magnetic field –upper panel for the scattering geometry). The solid lines are Lorentzian distribution fits to the data with constant width at half maximum σ ~ 0.18° (similar fits can be obtained using Gaussian distributions). Those data show that the diffracted intensity (rocking curves) collapse without any broadening above ~0.7T.

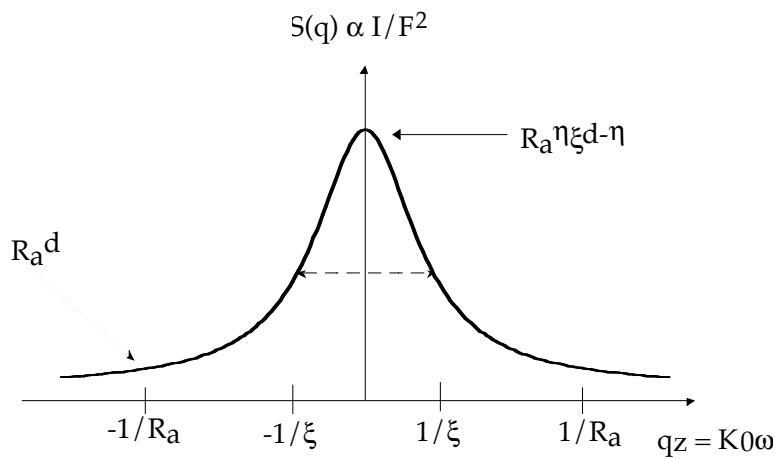

Figure 2: Bragg glass predictions for the angular dependence of the neutron diffracted intensity. In this phase the positional order is expected to decay with an exponent $\eta$ beyond some characteristic length scale $R_a$. $\xi$ is the experimental resolution. The arrows indicate the values of $S(q)$ for $q=0$ and $1/R_a$ respectively. If the disorder increases ($R_a$ decreases), the height of the peak decreases as $R_a^\eta$ but the peak does not broaden since the half width at half maximum is always given by the experimental resolution $1/\xi$ and not $1/R_a$. Such behavior can only be observed if $\eta \sim d$ (the dimensionality of the system) and is thus intrinsically related to the weak decay of the correlation functions. The height of the peaks gives a direct measure of the characteristic length $R_a$. To make the connection with Fig 1, one should set $d=1$ due to the integration of the diffracted intensity in the detector plane and $q$ should be replaced by $q_z$. As done in Fig 1, because $q_z = K_0 \omega$ an extra normalization factor $a_0$ is necessary to ensure that $K_0 \int I(\omega) d\omega = \int S(q) dq$ (note that the experimental resolution is constant in angle).

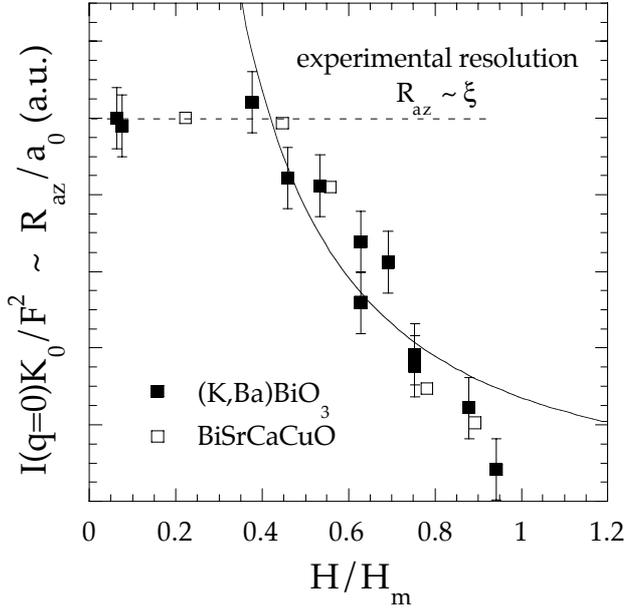

Figure 3: Magnetic field dependence of the positional length $R_{az}$ for $(K,Ba)BiO_3$ and BiSrCaCuO (ref 24) crystals. In the Bragg glass model $R_{az}$ is directly related the maximum of the diffracted intensity (intensity at the Bragg angle): $R_a \sim I(\omega=0)/F^2$. The x axis has been renormalized to the field $H_m$ above which the diffracted intensity disappears. The solid line corresponds to a $1/B^2$ dependence for $R_{az}$ as expected from a classical elastic theory. The horizontal dotted line is the limit above which $R_{az}$ becomes larger than the experimental resolution $\xi \sim 50 a_0$ in $(K,Ba)BiO_3$. The intensity drops rapidly towards zero at $H_m$ probably reflecting the proliferation of dislocations at the order-disorder transition

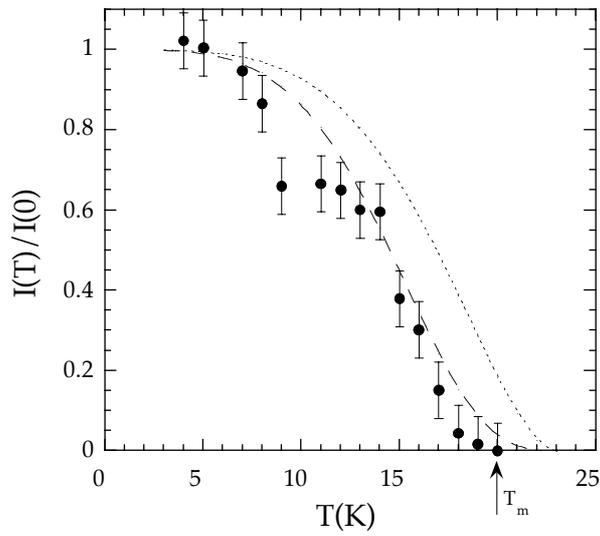

Figure 4: Temperature dependence of the maximum of the neutron diffracted intensity in a (K,Ba)BiO$_3$ crystal. The intensity (I) at the Bragg angle has been normalized to its value at T=0K. The dotted line is the $1/\lambda^4$ dependence expected in the London theory. As shown the data rapidly deviate from this simple dependence. The dashed line is the Bragg glass prediction (for $R_a < \xi$) for which I is expected to be proportional to $R_a(T)/\lambda^4(T)$. The order-disorder transition temperature can be obtained at 0.4T as I tends towards zero at $T_m \sim 20K$ (arrow).